\documentclass[11pt]{article}
\usepackage{moriond,epsfig}

\def\be{\begin{equation}}
\def\ee{\end{equation}}
\def\bea{\begin{eqnarray}}
\def\eea{\end{eqnarray}}

\newcommand{\fref}[1]{\mbox{{Fig.~}\ref{#1}}}

\newcommand{\sref}[1]{\mbox{{Sect.~}\ref{#1}}}
\begin{document}
\vspace*{4cm}
\title{A Study of the Charge of Leading Hadrons in Gluon and Quark
  Fragmentation} 

\author{ Martin Siebel}

\address{Fachbereich C, Bergische Universit\"at Wuppertal,\\
42097 Wuppertal, Germany}
%
%
\maketitle\abstracts{
In this study the electric charges of leading systems in quark and gluon jets
from hadronic three-jet events in $e^+e^-$-annihilation measured with the {\sc
  Delphi}-experiment are examined. Leading systems are defined by a rapidity gap between the
leading system of a jet and the rest of the event.
The measured charge distributions are compared with results from Monte-Carlo
simulations which do not contain colour-octet 
neutralisation processes. In the data
an enhanced production of neutral leading systems compared to Monte-Carlo
predictions is found in gluon jets, which is compatible with the expectations
from colour-octet neutralisation.  The quark jet sample is found in agreement
with the simulation.
}
\section{Introduction}
The existence of hadrons which contain gluons as valence particles is one of
the possible features of QCD which has not yet been experimentally
verified. The production of these particles requires a process in
which colour charges are balanced via colour-octet charges of gluons. 
This process should especially occur in gluon rich environments as
e.g.~in gluon jets from hadronic three-jet events in $e^+e^-$-annihilation.
However, this colour-octet neutralisation
process is expected to be obscured by higher order
 colour-triplet neutralisation processes. It has been suggested 
to investigate jets with a large gap in rapidity between the leading system
and the bulk of the jet \cite{suggest}. The presence of a rapidity gap is a
signature of an early colour decoupling of the leading system from the rest of
the event. Due to the early decoupling, higher order triplet processes are
suppressed and the effects of colour-octet neutralisation are expected to be
observable. 

Current fragmentation models only consider the neutralisation of
colour charges via the production of quark-antiquark pairs, i.e.~colour
triplet neutralisation. Due to the production of different quark flavours, the
leading system is allowed to have charges of +1, 0 and -1 in lowest
order. Colour octet neutralisation on the other hand only allows for neutral
leading systems, as gluons do not carry an electric charge. The expected
signature of colour-octet neutralisation is therefore a higher fraction of
neutral leading systems in gluon jets than it is predicted by Monte-Carlo
models which only account for colour-triplet neutralisation. 
\section{Data Analysis} \label{s:data}
The data analysed for this study\cite{wienpaper} have been recorded 
with the {\sc
  Delphi}-experiment at LEP in the years 1994-95 with a center-of-mass energy
  of $\sqrt{s}=91.2$ GeV. After applying cuts to select well-measured hadronic
  events, three-jet events have been selected using the
  \mbox{Durham-algorithm \cite{durham}} 
  with $y_{\mathrm{cut}}=0.015$. Additional
  cuts to provide clearly separated jets leave 314000 selected events. 

 Gluon and quark jets are identified in an implicit way assuming that the most
 energetic jet of an event (jet 1) is a quark jet and the least energetic jet
 (jet 3) is the gluon jet. This method provides purities $\ge90\%$ for the
 quark jet sample and $\sim 70\%$ for the gluon jet sample. Alternatively,
 gluon jets are identified for cross-check reasons using a tagging technique
 in events with initial $b$-quarks raising the gluon jet purity to 88\%. 
Details about the tagging procedure and
 the applied cuts are given elsewhere\cite{gluontag}.
The leading system of a jet is defined by the requirement that all charged
particles of the jet have a rapidity larger than a given cut-value $\Delta
y$. 
For central results $\Delta y=1.5$ is chosen.
The inclusion of neutral
particles 
in the rapidity gap definition 
does not affect the results of this analysis.

The production of rapidity gaps is suppressed more strongly in the
fragmentation of gluons than in the fragmentation of quarks. This leads to a
depletion of the gluon jet contribution to a given jet sample when a rapidity
gap is demanded. The purity of the gluon jet sample in dependence of the size
of the rapidity gap has been obtained from Monte-Carlo
simulation. Additionally, the gluon jet purity has been deduced from the
reduction rate of the three explicitely tagged jet samples (gluon jets,
$b$-jets, untagged) when a rapidity gap is demanded.
The gluon jet
purity with a rapidity gap of $\Delta y=1,5$ 
taken from Monte-Carlo (calculated from the reduction rate) is 47.2\% (47.4\%)
for the gluon jet sample defined by energy ordering and 82.0\% (80.4\%) for
the explicitely tagged gluon jet sample. The purities obtained with both
methods 
are in good agreement. 

The data are compared to three different Monte-Carlo models: {\sc
  Jetset}\cite{jetset}  with
and without simulated Bose-Einstein correlation (BEC) and {\sc
  Ariadne}\cite{ariadne} without BEC. Mesons of the same charge are pulled
  closer together due to the Bose-Einstein correlation. This can affect the
  charge distribution of the leading systems. As the magnitude of this
  disturbance is not
  known, the Monte-Carlo model with BEC is taken into account to study
  systematic effects. Differences in
  the deviation from the Monte-Carlo models are included in the systematic
  error of this study. The generated events are processed with a full
  simulation of the {\sc Delphi}-detector and the analysis chain described
  above before they are compared to the data.
\section{Results} \label{s:results}
\begin{figure}
\begin{center}
\epsfig{file=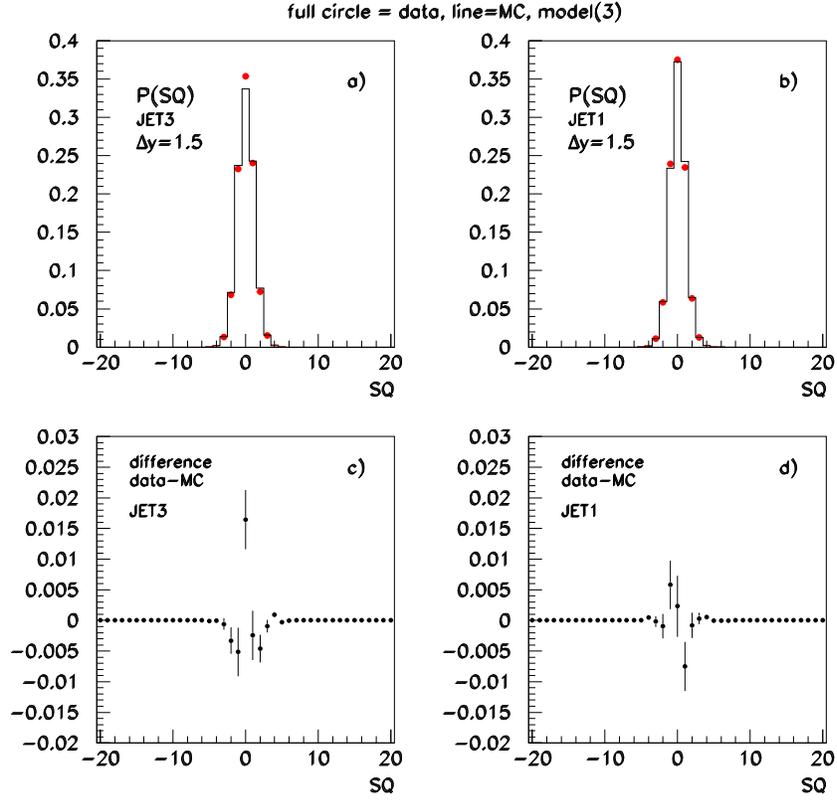,width=11cm}
\end{center}
\caption{{\bf Top:} 
   The distributions of the charges of leading systems in gluon
  (left) and quark (right) jets compared to the predictions of the {\sc
  Ariadne} Monte-Carlo model (solid lines).
  {\bf Bottom:} The difference between data and Monte-Carlo.
\label{fig:chdis}}
\end{figure}
\begin{figure}
\begin{center}
\epsfig{file=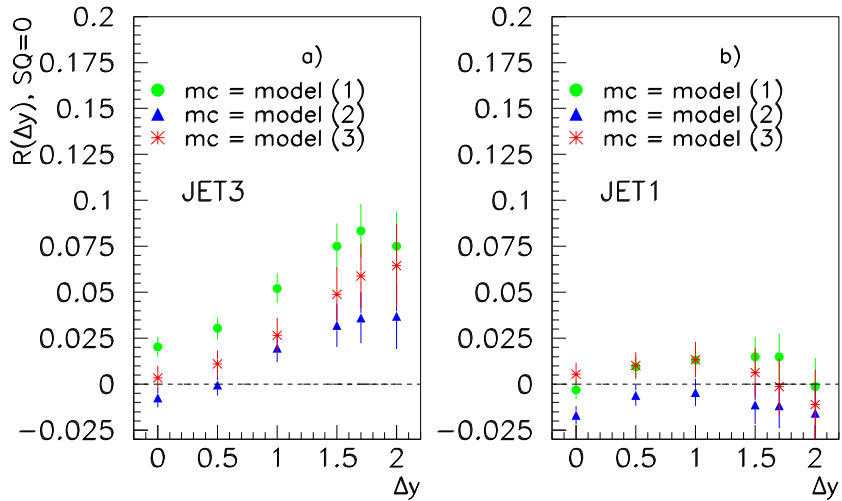,width=12.5cm}
\end{center}
\caption{The dependence of the relative difference between data and
  Monte-Carlo simulation in gluon jets (left) and quark jets (right) 
  as a function of $\Delta y$ for the
  three Monte-Carlo models {\sc Jetset} with BEC (model 1), {\sc Jetset} 
  without BEC (model 2) and {\sc Ariadne} without BEC (model 3).
\label{fig:ydep}}
\end{figure}
In \fref{fig:chdis} the distributions of the sum of charges ($SQ$) in 
the leading systems in gluon
jets (left) and quark jets (right) with a rapidity gap of $\Delta y=1.5$ 
are shown. The jets are identified by
energy ordering. The solid histograms indicate the distributions obtained from
the {\sc Ariadne} Monte-Carlo simulation. The distribution for quark jets is
well described by the Monte-Carlo simulation while for gluon jets the
occurrence of neutral leading systems is underestimated by the simulation. This
is the expected behaviour, if colour-octet neutralisation is present in the
data. In the two lower plots of \fref{fig:chdis} the difference between the
$SQ$-distributions from data and Monte-Carlo are shown. The surplus of neutral
leading systems in gluon jets is an effect of $\sim 3 \sigma$, while the
difference seen in the quark jet sample is compatible with zero. 

In order to
study the dependence of the effect on the chosen rapidity gap size, the
relative deviation $R(\Delta
y)=(P_{\mathrm{data}}(SQ=0)-P_{\mathrm{MC}}(SQ=0))/P_{\mathrm{MC}}(SQ=0)$
between data and simulation is studied for several values of
$\Delta y$. In \fref{fig:ydep} $R(\Delta y)$ is shown for all three used
simulations using jets identified by energy ordering. 
A nearly linear increase of $R$ with $\Delta y$ can
be observed for gluon jets while $R$ stays constant and compatible with zero
for quark jets.  However, the Monte-Carlo model including BEC
shows a non-vanishing value for $R$ in gluon jets
also for $\Delta y=0$.
The slope of $R(\Delta y)$ is
roughly the same for all three models studied. The discrepancy introduced due
to BEC seems therefore to be independent of of the rapidity gap size. 
In order to eliminate this influence of the BEC on this study, the variable
$R'(\Delta y) = R(\Delta y) - R(0)$ is studied, where the difference obtained
with a given simulation at $\Delta y=0$ is subtracted from the $R$-values
obtained with this simulation. The $R'$ values obtained for all three
simulations are in good agreement for a given $\Delta y$, 
remaining differences are added to the systematic error.

The results obtained with the explicitely tagged gluon jet sample are
consistent with the observations described above. Due to the higher gluon
purity, the effect is roughly two times the size of the effect in the energy
ordered gluon jet 
sample as it is expected from colour-octet neutralisation. However, due to the
limited statistics of tagged gluon jets, the statistical error increases,
leaving the effect less significant in this sample than in the gluon jet
sample obtained by energy ordering. Using the gluon jet sample
purities obtained as described in \sref{s:data}, the size of the effect in a
pure gluon jet sample can be calculated. The differences of $R'$ values and
estimated gluon purities between the 3 models 
are included in the systematic
error giving 
\be
R'(\Delta y=1.5) = 0.10 \pm 0.02_{\mathrm{stat.}}\pm 0.03_{\mathrm{syst.}}
\ee
for a pure gluon jet sample. While a variation of the event reconstruction
quality and the track finding efficiency have no significant effect on the
result, a variation of the parameters of the Monte-Carlo simulation, where
different sets of tuned parameters have been used, leads to an uncertainty of
$\pm 0.025$ on $R'$, which is included in the given systematic uncertainty. In
order to check, if the good agreement between data and Monte-Carlo found in the
energy ordered quark jet sample is mainly due to very hard tracks in the
leading jet, only tracks with $p\le30$ GeV have been taken into account with
no effect on the observed agreement.

The overproduction of neutral leading systems is only observed in gluon
jets, the effect increases with higher gluon jet purity and with increasing
rapidity gap size. All findings are in agreement with the expectations for
colour-octet neutralisation in gluon jets.
\section*{Acknowledgements}
I would like to express my gratitude to 
the Marie-Curie programme of the European Union whose
grant enabled me to take part in this conference. I also would like to
thank 
B.~Buschbeck and F.~Mandl for their support in the
preparation of this talk. 


\section*{References}
\begin{thebibliography}{99}
\bibitem{suggest} P.~Minkowski and W.~Ochs, Phys.~Lett.~{\bf B485} (2000) 139\\
P.~Minkowski and W.~Ochs, Proceedings 30th Int. Symp. on
  Multiparticle Dynamics, eds.~R.~Csorgo et al.~(WSPCE, Singapore 2001)
\bibitem{wienpaper} B.~Buschbeck and F.~Mandl, Proceedings 31st Int. Symp. on
  Multiparticle Dynamics, eds.~Bai Yting et al.~(WSPCE, Singapore 2002,
  p.50)
\\
  B.~Buschbeck and F.~Mandl, DELPHI-Note 2002-053-CONF-587, contributed paper
  to ICHEP 2002, Amsterdam
\bibitem{durham}
    S.~Catani et al., 
    Phys.~Lett. {\bf B269} (1991) 432
\bibitem{gluontag} {\sc Delphi}-collaboration, P.~Abreu et al.,
  Eur.~Phys.~J. {\bf C13} (2000) 573
\bibitem{jetset}
    T.~Sj\"ostrand,
    Computer Physics Comm. {\bf 82} (1994) 74
\bibitem{ariadne} L.~Loennblad, Computer Physics Comm. {\bf 71} (1992) 15
\end{thebibliography}
\end{document}